\documentclass[preprint,12pt]{elsarticle}

\usepackage{amssymb}
\usepackage{enumitem}
\usepackage{amsthm}
\usepackage{multirow}
\usepackage{amsmath,bm,amssymb}
\usepackage{url}

\journal{Computer Physics Communications}

\begin{document}

\begin{frontmatter}



\title{Particle Merging Algorithm for PIC Codes}


\author[label1]{M. Vranic}
\author[label1]{T. Grismayer}
\author[label1]{J. L. Martins}
\author[label1,label2]{R. A. Fonseca}
\author[label1]{L. O. Silva}
\address[label1]{GoLP/Instituto de Plasmas e Fus\~ao Nuclear, Instituto Superior T\'ecnico, Universidade de Lisboa, 1049-001 Lisbon, Portugal}
\address[label2]{DCTI/ISCTE - Instituto Universit\'ario de Lisboa, 1649-026 Lisboa, Portugal}

\begin{abstract}

Particle-in-cell merging algorithms aim to resample dynamically the six-dimensional phase space occupied by particles without distorting substantially the physical description of the system. Whereas various approaches have been proposed in previous works, none of them seemed to be able to conserve fully charge, momentum, energy and their associated distributions. We describe here an alternative algorithm based on the coalescence of N massive or massless particles, considered to be close enough in phase space, into two new macro-particles. The local conservation of charge, momentum and energy are ensured by the resolution of a system of scalar equations. Various simulation comparisons have been carried out with and without the merging algorithm, from classical plasma physics problems to extreme scenarios where quantum electrodynamics is taken into account, showing in addition to the conservation of local quantities, the good reproducibility of the particle distributions. In case where the number of particles ought to increase exponentially in the simulation box, the dynamical merging permits a considerable speedup, and significant memory savings that otherwise would make the simulations impossible to perform.

\end{abstract}

\begin{keyword}

partice-in-cell \sep coalescence scheme \sep QED cascade 


\end{keyword}

\end{frontmatter}

\section{Introduction}

Particle-in-cell (PIC) codes are a powerful tool of computational physics that allows to simulate non-linear evolution of  electromagnetic systems. The standard electromagnetic PIC algorithm relies on solving the relativistic Maxwell equations for the evolution of the fields, coupled with the relativistic Lorentz force to advance the charge density \cite{book_simulation}. This is a fully self-consistent model that starts from first principles and conserves the energy and momenta throughout the simulations (in fact PIC codes are either momentum conserving or energy conserving and no algorithm conserves both exactly). The particles can explore the full 6D phase space, while the fields are confined on a grid. Maxwell equations are solved at grid points, from where the fields later can be interpolated to any particle locations. Plasma particles are represented by a distribution of macro particles, that may carry different statistical weights (one macro particle can represent several real particles). Extended PIC codes can include ionization \cite{Chen2013, Bruhwiler}, binary collisions \cite{CollisionPeano, Takizuka} or quantum electrodynamics (QED) modules \cite{Timokhin, Nerush, Gremillet, Ridgers_solid,ThomasQED}. These codes have the capability to take full advantage of world's leading high-performance parallel computing systems - for example, the OSIRIS framework \cite{OSIRIS} has been shown to run efficiently on systems with as many as $10^5-10^6$ cores \cite{Fonseca_scaling}. The scalability relies on carefully optimised parallelisation that divides the space in a way that minimises communications and maximises load balance. 

While these codes have been successfully applied to a number of plasma physics scenarios, there are situations that are extremely difficult to model due to a significant accumulation of particles in a limited region of simulation space. For example, in QED cascades very localised regions of extremely strong field can easily produce vast numbers of electron-positron pairs even starting from just one seed electron, leading to an exponential growth of the number of particles being modelled, severely hindering simulation performance and eventually running out of memory. In principle, we could overcome this difficulty by resampling the 6D phase space with different macro-particles - many original macro-particles can be merged into fewer macro-particles with higher statistical weights. This is critical for simulating plasma in extreme conditions. However, one needs to ensure that merging does not alter the physics, so a special care should be taken to preserve not only fundamental properties of the system, but also the local particle phase space distribution. 
 
Previous attempts to merge particles for QED cascades were focused on conservation of total quantities: Timokhin in ref.\cite{Timokhin} presented a simple scheme where excess particles are deleted and their statistical weight is redistributed evenly among the rest of the simulation particles. This conserves the total charge, but does not conserve total energy and total momentum. None of the quantities are conserved locally, and this introduces differences in the particle distribution. The authors in ref. \cite{Nerush} use a similar algorithm where the randomly selected particles are deleted while the charge, mass, and energy of the rest particles are increased by the charge, mass, and energy of the deleted particles, respectively. In refs. \cite{Lapenta_old, Lapenta_new} the authors present several coalescence and splitting schemes, but neither of them conserves the particle distribution function both locally and globally. 

In this paper we present a different particle merging scheme that preserves the energy, momentum and charge locally and thereby minimises the potential influence to the relevant physics. The algorithm is applicable for massive particles (e.g. electrons, protons, positrons) or massless particles (photons). In addition, the algorithm naturally favourites faster merging in regions with many particles that have similar properties, and does not alter the tail of the distribution that is already sampled by only a small number of particles. All the particles that are merged together are close in 6D phase space. The main benefit of this scheme is that it allows for simulating scenarios that would otherwise be unaccessible, but it can also be used to accelerate simulations with high parallel load imbalance \cite{Fonseca_scaling} that can occur by accumulation of a large number of particles in a small region of space. 

The remainder of this paper is organised as follows: in section \ref{sec:algorithm} we describe the merging algorithm that conserves particle phase space distribution. In section \ref{sec:mergerate}, the theoretical estimates for the merging rate in the simulations are presented. Section \ref{sec:validation} focuses on the validation of the algorithm through examples of classical and QED plasma interactions: two-stream and filamentation instability, magnetic showers and QED cascades with a two-laser setup. Finally, we present our conclusions in section \ref{sec:conclusions}. 

\section{Algorithm}\label{sec:algorithm}

The goal of this algorithm is to map the coordinate and momentum phase space occupied by simulation particles, and resample it without changing the relevant properties of the particle distribution. This can be achieved by identifying the particles that are ``close'' to each other in 6D phase space (simultaneously close in coordinate and momentum space). The criteria on which particles are considered  ``close enough'' will depend on the typical length/momentum scales that appear in a specific physical scenario. Here we consider the general problem and then, for each specific example, we address the criteria to determine some of the key parameters of the algorithm (merging rate, sampling of the phase space).

\begin{figure}
\centering
\includegraphics[width=1.0\textwidth]{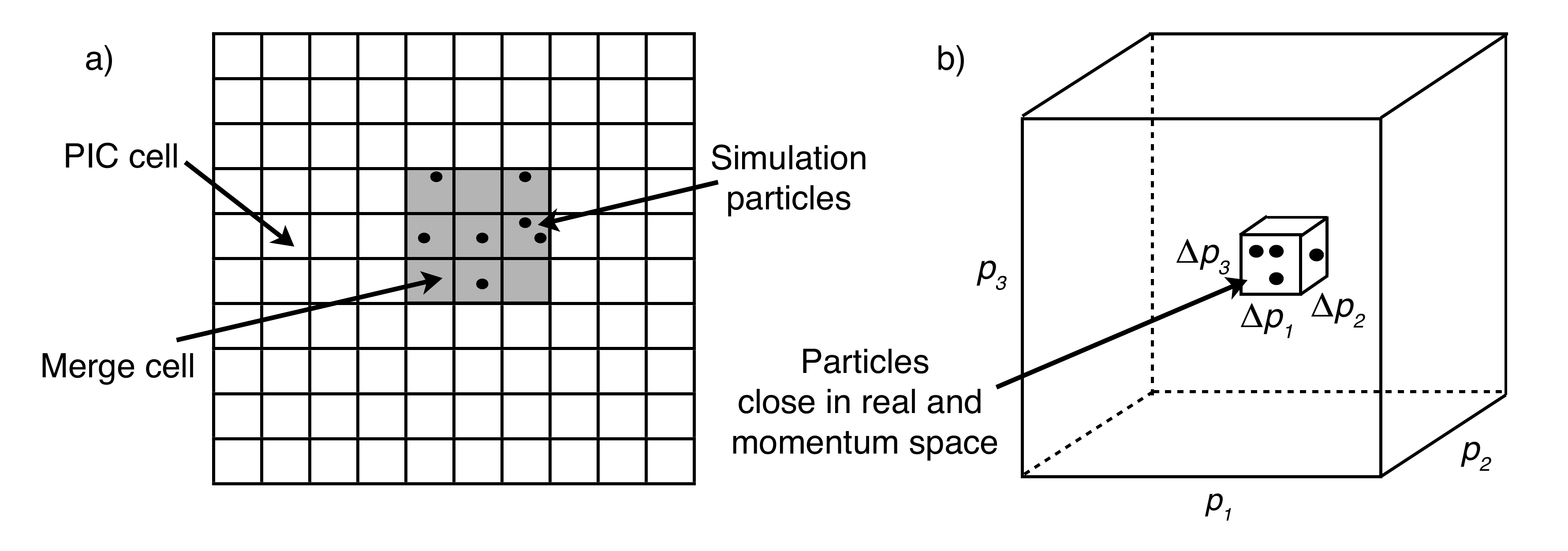}
\caption{Phase space mapping for the merging algorithm. a) An example of a merge cell in a 2D spatial grid. b) Momentum space within a single spatial merge cell. The small sub-cube represents a momentum cell, within which the particles are merged.} 
\label{merge_space}
\end{figure}

For now, we consider that a spatial merge cell contains an integer number of PIC cells in each direction, to be defined for each problem. Our division of space is shown in fig. \ref{merge_space} a) on an example of a 2D grid. Here, a merge cell is shaded and contains 9 PIC cells (3$\times$3).

For the particles that lie within a given merge cell, we first identify what are the boundaries of the momentum space ($p_{min}$ and $p_{max}$ in each direction of the momentum space). The 3D momentum space for merging is represented in Fig. \ref{merge_space} b) where it spans between the minimal and maximal momenta in each direction. Then, we divide this momentum space in several sectors per direction, which yields $n_1\times n_2 \times n_3$ volume elements that we define as the momentum cells. Currently all the momentum cells are uniformly distributed but the algorithm can be easily generalised for heterogeneously sized momentum cells. The particles that are within the same momentum cell (they are already in the same spatial merge cell) are considered to be close to one another in 6D phase space and, therefore, candidates to be merged together. 

It is now necessary to compute the total statistical weight $w_t$, momentum $\vec{p}_t$ and energy $\epsilon_t$ contained within one momentum cell: 
\begin{equation}\label{totalwpe}
w_{t}=\sum_{i=1}^N w_i\ , \quad \vec{p}_{t}=\sum_{i=1}^N w_i \vec{p}_i\ , \quad \epsilon_t=\sum_{i=1}^N w_i \epsilon_i\ .
\end{equation}
where $N$  is the total number of particles of the species to be merged within the momentum cell, while  $w_i$, $\vec{p}_i$ and $\epsilon_i$ represent the statistical weight, the momentum and the energy of the $i$-th particle respectively. For further calculations we introduce a normalised system of units: $p\rightarrow p/mc$, $\epsilon\rightarrow \epsilon/(mc^2)$, $t\rightarrow t\omega_N$, $E\rightarrow eE /(mc\omega_N) $,  $B\rightarrow eB /(mc\omega_N) $, where $c$ is the speed of light, $m$ is  the electron mass, $e$ elementary charge and $\omega_N$ a normalising frequency (typically it is equal to the background plasma frequency or the laser frequency). All total quantities defined in Eq. \eqref{totalwpe} should be conserved after merging the particles. Ideally, one would conceive that the merging process would lead to one macro particle per momentum cell; however, this does not allow to conserve all the significant quantities. Let us assume that there exists a particle that would conserve $w_t$, $\vec{p}_t$ and $\epsilon_t$. The weight $w_n$, momentum $\vec{p}_n$ and energy $\epsilon_n$ of such new particle would then be: 
\begin{equation}\label{single_part}
w_n=w_t\ , \vec{p}_n=\frac{\vec{p}_t}{w_t}\ , \quad \epsilon_n=\frac{\epsilon_t}{w_t}
\end{equation}
Such a particle would also need to satisfy an energy-momentum relation (in normalised units,  for electrons it takes the form $\epsilon_n^2=||\vec{p_n}||^2+1$,  and for photons $\epsilon_n=||\vec{p_n}||$). A simple example that illustrates a scenario where this is not satisfied is when initially we have only two particles in the momentum cell that have exactly the same weight $w$ and energy $\epsilon$, but opposite non-zero momentum vectors $\vec{p}$ and $-\vec{p}$. Here, $\vec{p}_t=0$ leading to also $\vec{p}_n=0$, $w_n=w_t=2w$ and $\epsilon_t=2w\epsilon$ leading to $\epsilon_n=\epsilon$. If the particles to be merged are photons, the energy-momentum relation is not valid for the new particle because $\epsilon=||\vec{p}||>0$ so $\epsilon_n>||\vec{p_n}||=0$. Similarly, for electrons $\epsilon=\sqrt{||\vec{p}||^2+1}>1$, hence $\epsilon_n>\sqrt{||\vec{p_n}||+1}=1$.

The previous example shows that merging into one macro particle would not always allow to locally conserve all the quantities we are interested in, as expected from the requirement to simultaneously conserve momentum and energy i.e. elastic merging. However, if the merging process results in two macro particles instead of one, all the relevant conservation laws can be satisfied. Let us consider two macro particles $a$ and $b$ with $w_a$, $\vec{p}_a$, $\epsilon_a$ and $w_b$, $\vec{p}_b$, $\epsilon_b$. To conserve the weight, momentum and energy they have to satisfy the following relations: 
\begin{align}
w_t&=w_a+w_b\ ,\nonumber \\  
\label{conservation} \vec{p}_t & =w_a \vec{p}_a + w_b \vec{p}_b \ , \\ 
 \epsilon_t&=w_a \epsilon_a + w_b \epsilon_b  \ . \nonumber  
\end{align}
Besides eqs. \eqref{conservation}, there are two more energy-momentum relations to be satisfied  
\begin{align} 
\label{photmomene}&\mathrm{for\  photons~(massless~particles)}:&&\  \epsilon_a = p_a \ , \quad \epsilon_b = p_b\ ; \\
\label{elemomene}&\mathrm{and\ for\ electrons~(massive~particles)}:&&\   \epsilon_a^2 = p_a^2+1 \ , \quad  \epsilon_b^2 = p_b^2+1 \ . 
\end{align}
From now on, we will consider, without loss of generality the massive particles to be electrons, but the algorithm is valid for other massive particles as well. Equations \eqref{conservation}, \eqref{photmomene} or \eqref{elemomene} make for a system of 7 scalar equations to be satisfied by the proper choice of 10 scalar variables. For the sake of simplicity, we assume that the merged particles are identical i.e. $w_a=w_b=w_t/2$ and that $\epsilon_a=\epsilon_b=\epsilon_t/w_t$. From \eqref{conservation} we then get 
\begin{equation}\label{mom}
\vec{p}_a + \vec{p}_b = \frac{2\vec{p}_t}{w_t}.
\end{equation}
From relations \eqref{photmomene} and \eqref{elemomene} we can express $p_a=p_b=f(\epsilon_t/w_t)$. We will not express it explicitly so that we can continue explaining the algorithm without making the choice if our particles are photons or electrons. For now, we assume that we can calculate the value of $p_a$ and that $p_a\geq p_t/w_t$ (the inequality follows from geometry and will be proven later).  
 
\begin{figure}
\centering
\includegraphics[width=0.9\textwidth]{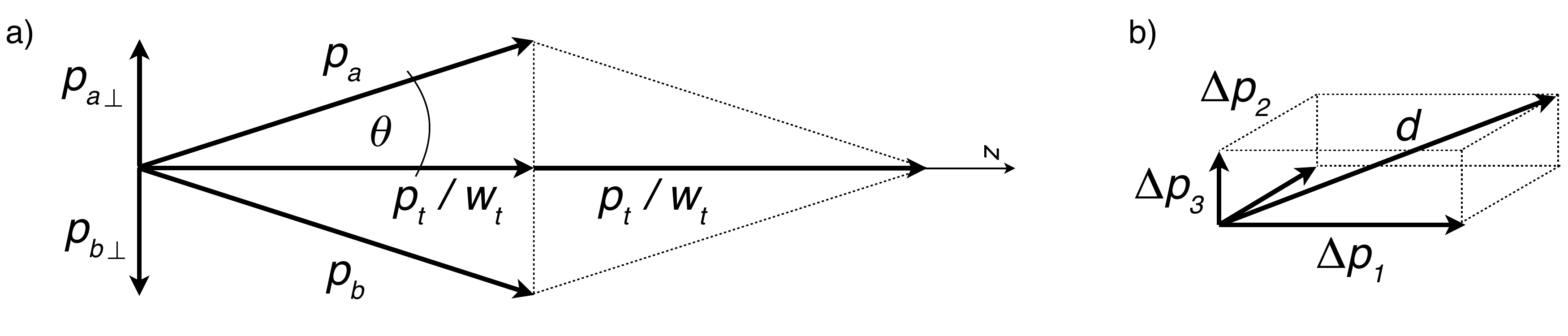}
\caption{a) Planar view of the two new particles momentum vectors $\vec{p}_a$ and $\vec{p}_b$ that make $\vec{p}_a + \vec{p}_b =2 \vec{p}_t/w_t$. b) Diagonal vector of the momentum cell $\vec{d}=(\pm \Delta{p}_1,\ \pm \Delta{p}_2,\ \pm \Delta {p}_3)$. } 
\label{merge_vectors}
\end{figure}

Figure \ref{merge_vectors} shows a plane that contains the direction of $\vec{p}_t$ and illustrates how two particles can satisfy Eq. \eqref{mom}. It is enough that they have same momentum components parallel to the total momentum $(\vec{p}_{a})_{||}=(\vec{p}_{b})_{||}=\vec{p}_t/w_t$ and the same  magnitude of the antiparallel components perpendicular to the total momentum $(\vec{p}_{a})_{\perp}=-(\vec{p}_{b})_{\perp}$. The angle $\theta$ between $\vec{p}_a$ and the direction of the total momentum $\vec{p}_t$ is determined by 
\begin{equation}\label{cosinus}
\cos \theta = \frac{p_t}{w_t p_a}\ ,
\end{equation}
and $(p_a)_\perp=p_a \sin \theta$. If we choose a spherical coordinate system ($r,~\theta,~\phi$) with the $z$-axis in direction of $\vec{p}_t$, it is clear that there is an infinite number of vectors that satisfy Eq. \eqref{mom}. In fact, from the previous considerations, there is still an arbitrary variable, and we could choose an arbitrary azimuthal angle $\phi$ for the vector $\vec{p}_a$ as long as it makes angle $\theta$ with the $z$-axis, still satisfying Eq. \eqref{mom}. Once $\vec{p}_a$ is chosen, this determines $\vec{p}_b$ as well. Particle momenta chosen in this algorithm obey all the necessary constraints and conserve the weight, energy and momentum locally. 

Even though $\phi$ and the plane in Fig. \ref{merge_vectors} a) can be chosen arbitrarily, while simultaneously guaranteeing the total momentum within the momentum cell is conserved, we note that this arbitrariness could distort the final distribution function. Let us assume a plasma that does not move in the $x_3$ direction ($p_3=0$), but has a very large momentum spread in other 2 directions. The $\vec{p}_t$ of any momentum cell within any merging cell will be confined in the $p_1$-$p_2$ plane, but if we choose the plane of vectors $\vec{p}_a$ and $\vec{p}_b$ arbitrarily, this may result in a resulting merged particle having a nonzero component in the $x_3$ direction. 
To avoid such effects, we should make the choice of a plane such that it naturally favours the momentum spreading of the merged particles $\vec{p}_a$ and $\vec{p}_b$ in the direction where the momentum spread already exists within the momentum cell. 

To do so, we can use one of the space diagonal vectors joining the vertices of the momentum cell to form the plane with $\vec{p}_t$ (see Fig. \ref{merge_vectors} b)). This immediately guarantees that if there is no motion in one of the directions, merging will not introduce any spreading along that direction (i.e. if both $\vec{p}_t$ and $\vec{d}$ are in $x_1$-$x_2$ plane, the result will also be in $x_1$-$x_2$ plane). If the diagonal chosen is collinear with $\vec{p}_t$, we can specify another diagonal, provided that there are at least two directions where the momentum spread is non-zero. Special care should be taken when both momentum and momentum spread exist in one direction only - an example would be a particle beam with finite energy spread in $p_1$ moving in $x_1$ direction. We note that for photons, this is easily solved, because then $\epsilon_t=p_t$ and it is even possible to initialise one photon instead of two, while still conserving all the main quantities. For electrons, this can not be guaranteed and electrons in these conditions should not be merged. 

After we have decided what are the momenta of the two new particles within the momentum cell, what is left is to decide where these particles will be initialised. It seems natural to arrange these two particles in the vicinity of the centre-of-mass of the group of particles they are replacing. But, here we will recall that a merge cell can contain several PIC cells, and the centre-of-mass of a sample of particles within a merge cell is more likely to be located in the central PIC cells. This is true for all the momentum cells within the merge cell, so we may put many particles created by merging in a small area of the merge cell. Therefore, if we pick the positions as centre-of-mass positions, we may be introducing local spikes in the density. To avoid this, we pick randomly two already existing particles within the momentum cell and put the new particles exactly at their positions. In this way,  artificially induced spikes in the density will not appear provided that we have a large enough statistical sample, (which is automatically guaranteed because the merging is performed only when the number of particles in a merging cell becomes very large).  

\begin{figure}
\centering
\includegraphics[width=0.9\textwidth]{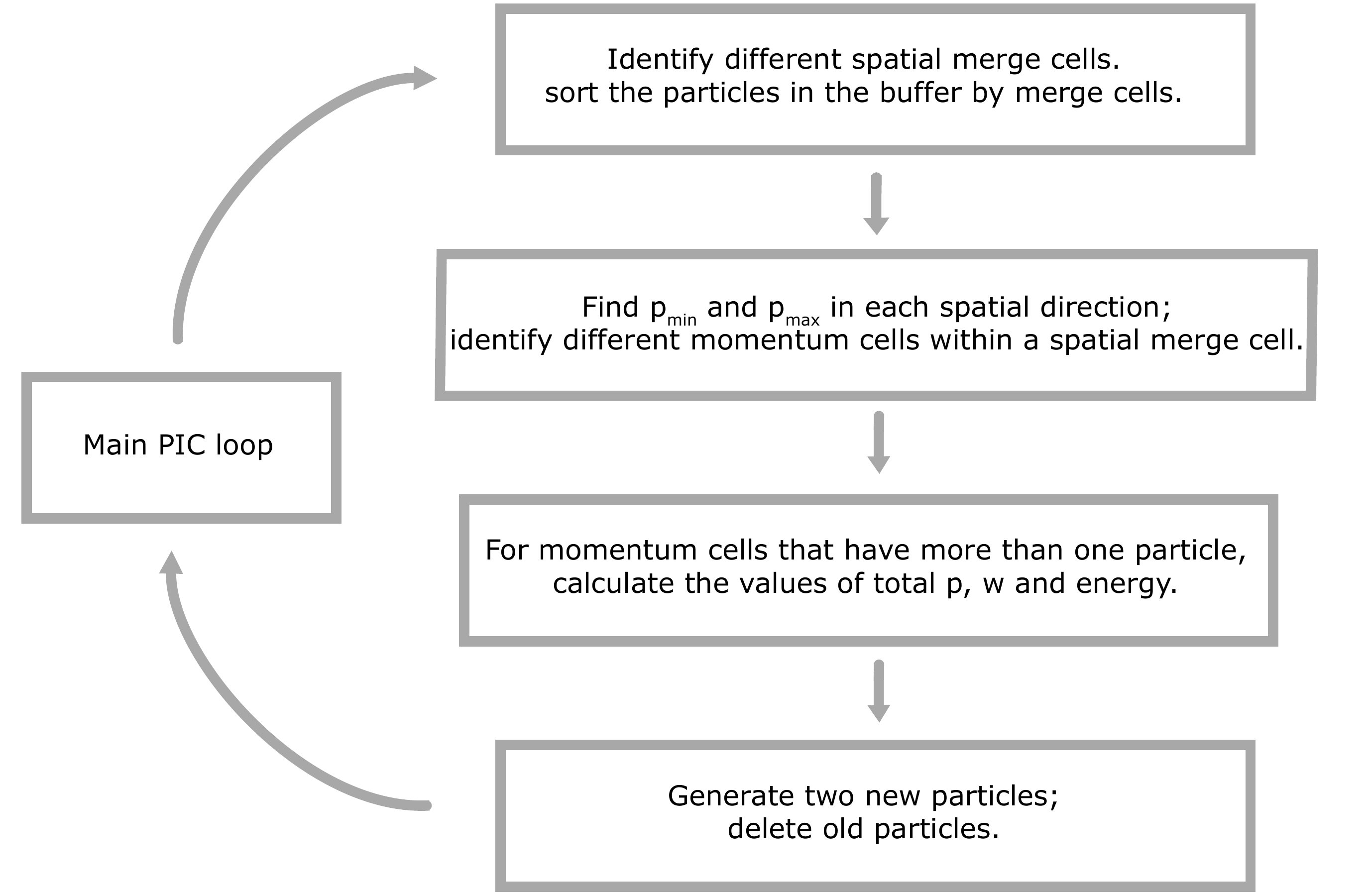}
\caption{Summarised loop of the merging algorithm.} 
\label{merge_loop}
\end{figure}

\subsection{Proof that $p_a\geq \frac{p_t}{w_t}$}
It is clear from Eq. \eqref{cosinus} that for $p_a<p_t w_t$ the above presented recipe would not give a sensible result (i. e. $\cos \theta > 1$), and, therefore, it is essential to show that the inequality $p_a\geq p_t/w_t$ is always true.     
For photons, $p_a=\epsilon_a=\epsilon_t/w_t$, so the inequality is equivalent to $\epsilon_t\geq p_t$ (the weights are always positive). In terms of a sum over the original photons this is written as
\begin{equation}\label{eq8}
\sum_i w_i \epsilon_i \geq \left|\sum_i w_i \vec{p}_i \right| ,
\end{equation}
where in the left-hand side we can use the relation $\epsilon_i=p_i$ and Eq. \eqref{eq8} can be re-written as
\begin{equation}\label{phot_trivia}
\sum_i w_i p_i \geq \left|\sum_i w_i \vec{p}_i \right| .
\end{equation}
The inequality \eqref{phot_trivia} is always satisfied for a set of vectors. This follows from the triangle inequality: the left-hand side is a fixed number and the right-hand side reaches the maximum when the vectors are all collinear and pointing in the same direction (then the sums are equal). 

For electrons, $p_a^2=\epsilon_a^2-1$. In this case, the inequality becomes 
\begin{equation}
\frac{\epsilon_t^2}{w_t^2}-1\geq \frac{p_t^2}{w_t^2}\ , \quad \mathrm{or} \quad \left( \sum_i w_i \epsilon_i \right)^2 \geq \left(\sum_i w_i p_i \right)^2 + \left(\sum_i w_i \right)^2 .
\end{equation}
Here, we already used the inequality  \eqref{phot_trivia} valid for any set of vectors, to obtain a fully-scalar sum. This then yields  
\begin{equation}
\sum_i w_i^2 \epsilon_i ^2 + \sum_{i, j,\  i\neq j} w_i w_j \epsilon_i \epsilon_j \geq \sum_i w_i^2 p_i ^2  + \sum_{i, j,\  i\neq j} w_i w_j p_i p_j + \sum_i w_i^2 + \sum_{i, j,\  i\neq j} w_i w_j 
\end{equation}
where we know that $\sum_i w_i^2 \epsilon_i ^2=\sum_i w_i^2 p_i ^2 +  \sum_i w_i^2$ because $ \epsilon_i ^2 = p_i ^2 + 1$ for every particle (every $i$). What is now left to prove now is 
\begin{equation}\label{ele_trivia}
 \sum_{i, j,\  i\neq j} w_i w_j \epsilon_i \epsilon_j \geq  \sum_{i, j,\  i\neq j} w_i w_j p_i p_j  + \sum_{i, j,\  i\neq j} w_i w_j \ .
\end{equation}
If, for every two particles ($i\neq j$), we demonstrate that $\epsilon_i \epsilon_j \geq p_i p_j + 1$, the inequality \eqref{ele_trivia} will automatically be satisfied as well. Expressing the energy through the energy-momentum relation once again yields: 
\begin{equation} 
\sqrt{p_i^2+1} \sqrt{p_j^2+1} \geq p_i p_j + 1.
\end{equation}
Since both sides are positive, we can square the inequality and obtain
\begin{equation}
p_i^2p_j^2+p_i^2+p_j^2+1 \geq p_i^2p_j^2 + 2p_i p_j +1
\end{equation}
which transforms to 
\begin{equation}
(p_i - p_j)^2\geq 0, 
\end{equation}
an inequality that is always satisfied. Therefore, for both photons and electrons $p_a\geq p_t/w_t$.

\section{Merging rate}
\label{sec:mergerate}

\begin{figure}[t!]
\centering
\includegraphics[width=1.0\textwidth]{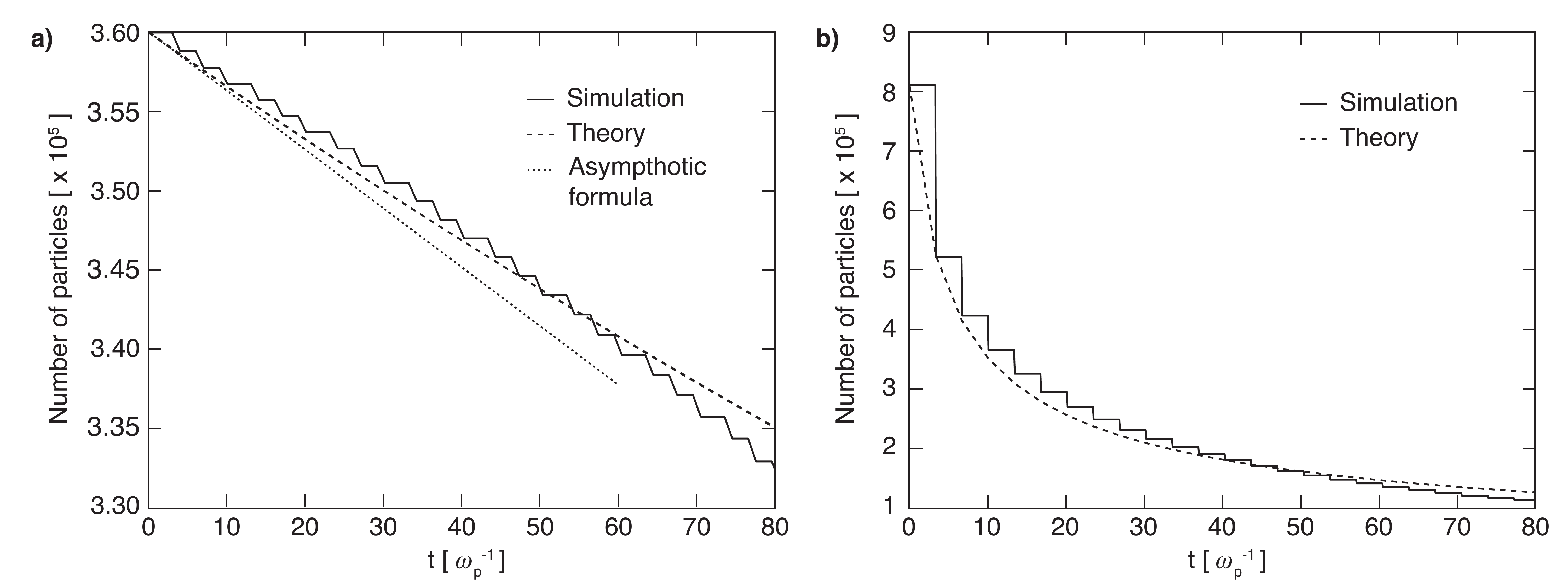}
\caption{ Number of particles as function of time for a 2D uniform thermal plasma. The initial velocity distribution is a waterbag distribution function in momentum space. The solid line shows the results of the PIC simulation, the dashed line represents the analytical prediction given by the numerical solution of Eq. (\ref{eq:mergrate}) and the dotted line represents the asymptotic solution (\ref{eq:asymp}),
for a) slow merging $\lambda = 0.144$ and b) fast merging $\lambda = 2.53$. } 
\label{merge_rate}
\end{figure}

The algorithm allows to merge, in a single momentum cell associated with a spatial merging cell, $N$ particles into two. Given all the momentum cells in the simulation, the number of particles $\Delta N_T$ that are removed from the simulation in a time interval $\Delta t_m$ (this time interval is problem dependent and corresponds to the inverse of the merging frequency, i.e., $\Delta t_m=1/\omega_m$) defines the merging rate of the algorithm. Determining the merging rate allows us to assess the impact and the efficiency of the algorithm in the overall evolution of the number of particles in the simulation. In the general case, the total number of particles $\Delta N_T$ being statistically deleted in a time $\Delta t_m$ is
\begin{equation}
\label{eq:Partdel}
\Delta N_T = \sum_{i=1}^{N_c}\sum_{j=1}^{N_m}\sum_{k=3}^{N_{p,i}}P_{ij}(k;N_{p,i},N_m)(k-2)
\end{equation}
where $N_c$ is the number of merging cells, $N_{p,i}$ is the number of particles in the $i$-th merging cell, $N_m = n_1\times n_2 \times n_3$ is the total number of momentum cells, and $P_{ij}(k)$ the probability of finding $k$ particles in the $j$-th momentum cell of the $i$-th merging cell. A rigorous calculation of $\Delta N_T$ is in general not possible since the distribution of particles in every merging cell is {\it a priori} unknown. However, we can choose a set-up such that Eq. (\ref{eq:Partdel}) simplifies drastically; the case of a uniform density thermal plasma with an initial waterbag momentum distribution offers the advantage of having an exact expression for the probability $P_{ij}(k)$. The uniform density implies that all merging cells should have almost the same number of particles. The same reasoning applies for the momentum space where the waterbag distribution ensures that the number of particles in a merging cell will be evenly distributed in all momentum cells. These properties are exact if the distribution is continuous. In the case of a discrete distribution, fluctuations arise due to the thermal motion of the particles. Let us now assume that we can neglect the fluctuations in the density so that the number of particle in each merging cell is considered as constant. Nonetheless, the statistical fluctuations associated to the distribution of the particles in the binned momentum space are of high relevance to compute the number of particles being merged. For a discrete uniform distribution (such as the waterbag distribution function), the probability of finding $k$ particles in a momentum cell (assuming that all momentum cells have here the same size) is the discrete Poisson probability: $P(k;\lambda) = \lambda^ke^{-\lambda} /k!$, where $\lambda = N_p/N_m$. Therefore the merging rate for a uniform thermal plasma is
\begin{equation}
\label{eq:mergrate}
\frac{dN_T}{dt}=-\omega_m N_cN_m\sum_{k=3}^{N_{p}}P(k;N_p/N_m)(k-2),
\end{equation}
where $\omega_m = 1/\Delta t_m$ is the merging frequency defined for each scenario. When the average number of particles per momentum cell is less than one, i.e., $N_p \ll N_m$, the parameter $\lambda$ is very small, the Poisson distribution reduces to $P(k;\lambda \ll 1) \simeq \lambda^k /k!$ and the result of the sum in Eq. (\ref{eq:mergrate}) comes mainly from the contribution of the first term of the sum, $P(k=3;N_p/N_m)$. The asymptotic formula, assuming $\lambda$ constant, for the merging rate reads 
\begin{equation}
\label{eq:asymp}
\frac{dN_T}{dt}\simeq-\omega_m \frac{N_cN_p^3}{6N_m^2}.
\end{equation}
\begin{table*}[t!]
\centering
\small

\begin{tabular}{|l||p{0.0mm}c|p{0mm}c|} 
\hline
\bf Test && \bf slow merging && \bf fast merging \\ 
\hline
\hline 
Dimension && 2D && 2D \\
Box size [$c/\omega_p$] && 5$\times$5 && 5$\times$5  \\
\# cells && 50$\times$50 && 50$\times$50  \\
$\Delta t\  [1/\omega_p]$ && 0.0672 && 0.0672  \\
\# Part/cell  && $12\times12$ && $18\times18$  \\
Merge frequency  && 50 $\Delta t$ && 50 $\Delta t$ \\
Merge cell size && 1$\times$1 && 2$\times$2 \\
Momentum cell && $10\times10\times10$ && $8\times8\times8$ \\
initial $\lambda = N_p/N_m$ && 0.144 && 2.53 \\
Thermal velocity && $v_x=v_y=v_z=0.1\ c $ && $v_x=v_y=v_z=0.1\ c $\\
\hline
\end{tabular}
\caption{Simulation parameters for the merging rate.}
\label{table:merge_rate}
\end{table*}

Surprisingly the merging persists for arbitrary small values of the ratio $N_p/N_m$, thus leading to the conclusion that the number of particle decreases linearly with time in the limit $N_p \ll N_m$. When the parameter $\lambda$ is not small compared to one, there is no simple expression for the merging rate and the Eq. (\ref{eq:mergrate}) should be evaluated numerically. 

To verify our predictions regarding the merging rate, we have performed simulations of thermal plasmas with initial waterbag distribution functions. Two simulations cases are presented here: a slow and a fast merging where the initial values of the parameter $\lambda$ were respectively chosen to be $\lambda = 0.144$ and $\lambda = 2.53$, corresponding to a finer (coarser) discretization of the momentum space begin resampled. The details of the two simulations can be found in the Table \ref{table:merge_rate}. The comparisons between the simulations and the Eqs.(\ref{eq:mergrate}) and (\ref{eq:asymp}) are shown in Fig. \ref{merge_rate}. For both cases an excellent agreement is found between simulation and theory. We observe that Eq. (\ref{eq:mergrate}) is only valid for uniform thermal plasmas with waterbag distribution function. Any deviations from this distribution would alter the predicted merging rate albeit keeping the same trend (if $\lambda \ll 1$, one expects a constant merging rate). For instance, a classical Maxwellian distribution spreads the particles in momentum space from approximatively $-5 p_{th}$ to $5 p_{th}$ (the particles out of this range represent a very small fraction of the total number, i.e., $1-\mathrm{erf}(5/\sqrt(2))$) whereas a waterbag distribution (corresponding to the same density and same amount of kinetic energy than the Maxwellian distribution) spreads exactly the particles from  $-\sqrt{3} p_{th}$ to $\sqrt{3} p_{th}$. Hence, for evenly spaced out momentum cells ranging from $p_{min}$ to $p_{max}$ in each direction, it is evident that a momentum cell for the Maxwellian will be bigger than for the waterbag distribution. Despite the different shapes of the two distributions, the merging rate corresponding to the Maxwellian distribution will be faster since the majority of the particles are actually contained into the bulk of the distribution. We have verified this in simulations with Maxwellian distribution functions.

\section{Numerical simulations}\label{sec:validation}

\begin{table*}[t!]
\centering
\small
\begin{tabular}{|l||p{0.0mm}c|p{0mm}c|p{0mm}c|p{0mm}c|} 
\hline
\bf Test &&\bf 2-stream && \bf Current && \bf Magnetic  && \bf QED  \\ 
\bf   && \bf   && \bf filamentation && \bf  shower && \bf  cascade \\ 

\hline
\hline
Dimensionality && quasi-1D && 2D &&2D && 2D\\
Norm freq [$\omega_N$]&& $\omega_p$ && $\omega_p$ && $\omega_c$=4.4$\times10^{14}/$s$$&& $\omega_0$=1.5$\times10^{15}/$s$$\\
Box size [$c/\omega_N$] && 10.0$\times$0.1 && 10.0$\times$10.0 && 2.0$\times$2.0 && 300$\times$120 \\
\# cells && 300$\times$5 && 100$\times$100 && 50$\times$50 && 3000$\times$1200 \\
$\Delta t\  [1/\omega_N]$ && 0.0163 && 0.0672 && 0.001 && 0.0692 \\
\# Part/cell  && $6\times6$ && $6\times6$ && $5\times5$ &&  $1\times1$ \\
Merge frequency  && 10 $\Delta t$ && 10 $\Delta t$ &&50 $\Delta t$ &&5 $\Delta t$ \\
Merge cell size && 1$\times$5 && 2$\times$2 && 1$\times$1 &&  10$\times$10 \\
Momentum cell && $20\times20\times20$ && $20\times20\times20$ && $20\times20\times1$ &&  $10\times10\times1$ \\
\hline
External field&& - && - && B$_3$=7.47$\times10^{10}$ G && $a_0$=1000 \\
Background && $e^+$ && $e^+e^-$ && - && $e^-$\\
Flows && $e^-e^-$ && $e^+e^-$ && $e^-$ && -\\
Flow velocity && $\pm 0.2\ c$ && $0.2\ c$ && $\gamma$ = 3000 && -\\
Thermal velocity && $0.001\ c$ && $0.001\ c$ && - && -\\
\hline
Run time wom&& 381 s && 139 s && 821 s && 29917 s\\
Run time wm&& 289 s && 103 s && 340 s && 1343 s\\
\hline
\# of Nodes && $2\times2$ && $2\times2$ && $2\times2$ && $100\times4$\\
\hline
\end{tabular}
\caption{Simulation parameters for benchmarks of the merging algorithm.}
\label{table:merge_param}
\end{table*}

We tested the merging algorithm in various scenarios to evaluate the effect of particle merging in the physics results. The problems tested ranged from classical plasma physics problems to more extreme scenarios that could be modeled without particle merging, showing excellent agreement between merged and non-merged simulations. We will focus here on 4 problems: i) 2 stream instability, ii) Current filamentation, iii) Magnetic shower and iv) QED cascade. Details for the simulation parameters can be found on table \ref{table:merge_param}. 

\subsection{Input parameters}

{\it Merging and momentum cells}: as we discussed in the algorithm description, the merging process needs to identify the particles that are close in the 6D phase space. The smallest possible merging cell, ensuring the closest proximity in real space, corresponds to a single simulation cell. However, we should keep in mind that an efficient merging is unlikely to occur in such scenario, given the small number of particles likely to be found in the merging cell. Larger merging cells should therefore be chosen, while ensuring that the merge cell size is still sufficiently smaller than the smallest relevant physical scale in the simulation. In the algorithm it is also necessary to specify the number of bins each momentum space is divided into. The test runs we have carried out seem to indicate that is preferable to use at least 8 bins per dimension.  This number may only be lowered in the dimensions where the momentum dispersion is absent.     

{\it Merging frequency}: the simulations performed with and without the merging algorithm show that the merging frequency cannot be chosen arbitrarily. Merging every at time step will tend to wash out some of the details of the microphysics, so the choice of the merging frequency  $\omega_m$ should be such that the smallest characteristic time scale $\omega_c$ of the system remains well described. The condition $\omega_m \sim \omega_c$ should therefore be sufficient to ensure that all relevant physics is accurately modelled. However this condition is not always applicable because it generally leads to a slow merging rate, and as a rule of thumb we recommend a lower limit for the merging frequency such that $1/\omega_m > 5 \Delta t$ with $\Delta t$ being the PIC time step. 

\subsection{Streaming instabilities}

\begin{figure}[t!]
\centering
\includegraphics[width=1.0\textwidth]{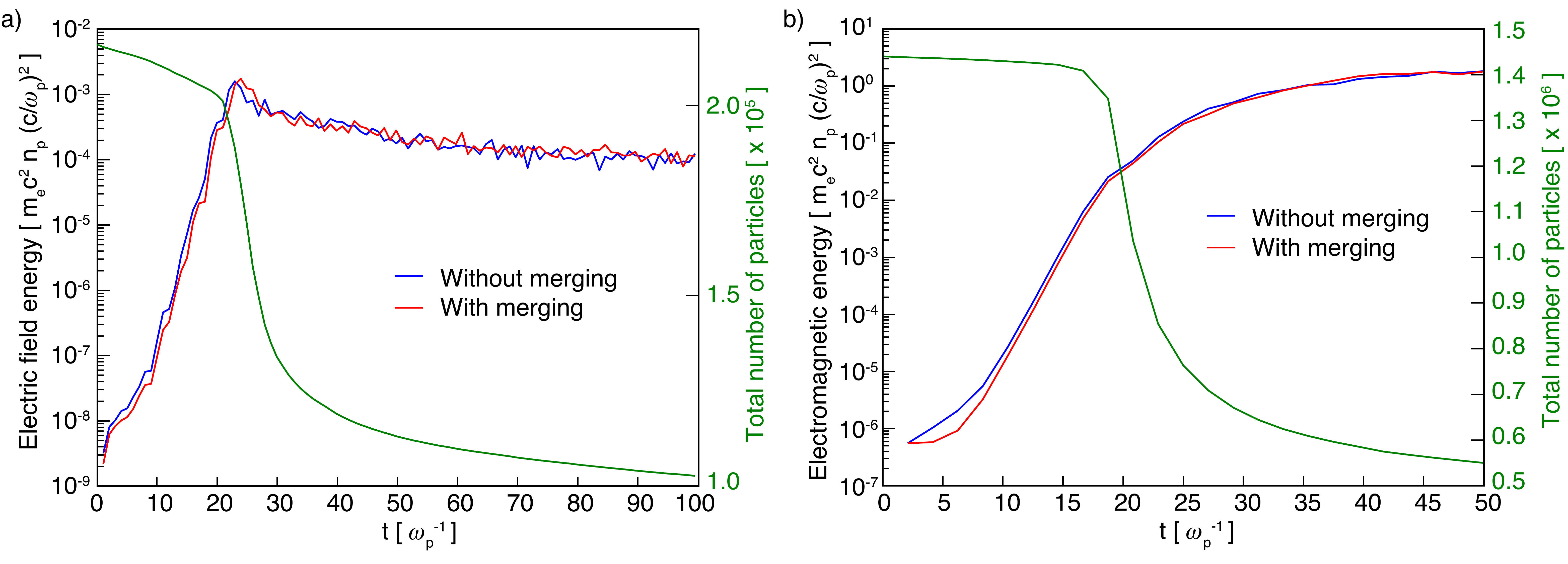}
\caption{a) Two stream instability. b) Current filamentation instability. About 50\% of particles are merged in both simulations. The left ordinate represent the electric/electromagnetic energy of the system while the right ordinate gives the total number of particle in the simulation. The blue and red lines depict respectively the total electromagnetic field energy of the system for the simulation with and without merging. The green lines show the total number of numerical particles in the simulation as a function of time. } 
\label{merge_stream}
\end{figure}

The cold two-stream instabilities, both electrostatic and purely electromagnetic (sometimes referenced as Weibel or current filamentation \cite{Medvedev_loeb}) have been studied by means of numerical simulations for decades \cite{Silva2003,Drummond,Bret,Thode_Sudan}. They represent simple setups that allow us to test validity of the merging algorithm that we have been describing in the previous sections. The simulations results for both instabilities are shown in Fig. \ref{merge_stream}. We observe an excellent agreement between the runs with and without merging, confirming that the algorithm does not alter the physics while merging the particles. As seen in Fig. \ref{merge_stream} the algorithm leads to a decrease of approximatively 50 \% of the total number of particles. We also see that both runs show a very similar trend: a slow merging rate during the linear phase of the instabilities (exponential growth of the field energy) followed by a fast merging during the early saturation and, finally, once again a slow merging after saturation. These three stages can be fully explained with our previous analysis of the merging rate. During the linear phase, $0< \omega_pt< 20$, the number of particles decreases approximatively linearly with time. This is because the linear phase consists of small perturbations originating from thermal noise, (that can be neglected as long as they remain very small compared to the zeroth order quantities). The plasma is hence still close to its initial zero-order equilibrium state and according to the predictions of section \ref{sec:mergerate}, the number of particles should decrease in a linear manner since the initial merging parameter $\lambda \ll 1$ for both simulations ($\lambda=6^2\times5/20^3=0.0225$ for the two-stream and $\lambda=6^2\times4/20^3=0.018$ for the current filamentation). The second stage is characterised by the saturation of the instabilities, corresponding to the time interval $20<\omega_pt<30$ in Fig. \ref{merge_stream}. At saturation, the linear perturbed fluid quantities start to be on the order of the equilibrium parameters and, in the case of the two-stream instabilities, the density of the plasma exhibits strong modulations that can reach several times the initial density. The bunching in configuration space is accompanied by a bunching in momentum space that allows more particles to be merged. The parameter $\lambda$ can reach values above one during this phase and one observes a fast merging with a curve that resembles the one obtained in section \ref{sec:mergerate}. Finally, the last stage is somehow similar to the first one. During the non-linear phase, 
the kinetic energy of the flow is converted into electromagnetic fields and thermal particles. The spikes of density are less pronounced as the time goes by and as a result the density tends to be more uniform. This is thus similar to the case of a thermal plasma with weakly modulated density which induces a slow merging, as seen in Fig. \ref{merge_stream}. 

\subsection{Magnetic showers}

\begin{figure}[t!]
\centering
\includegraphics[width=1.0\textwidth]{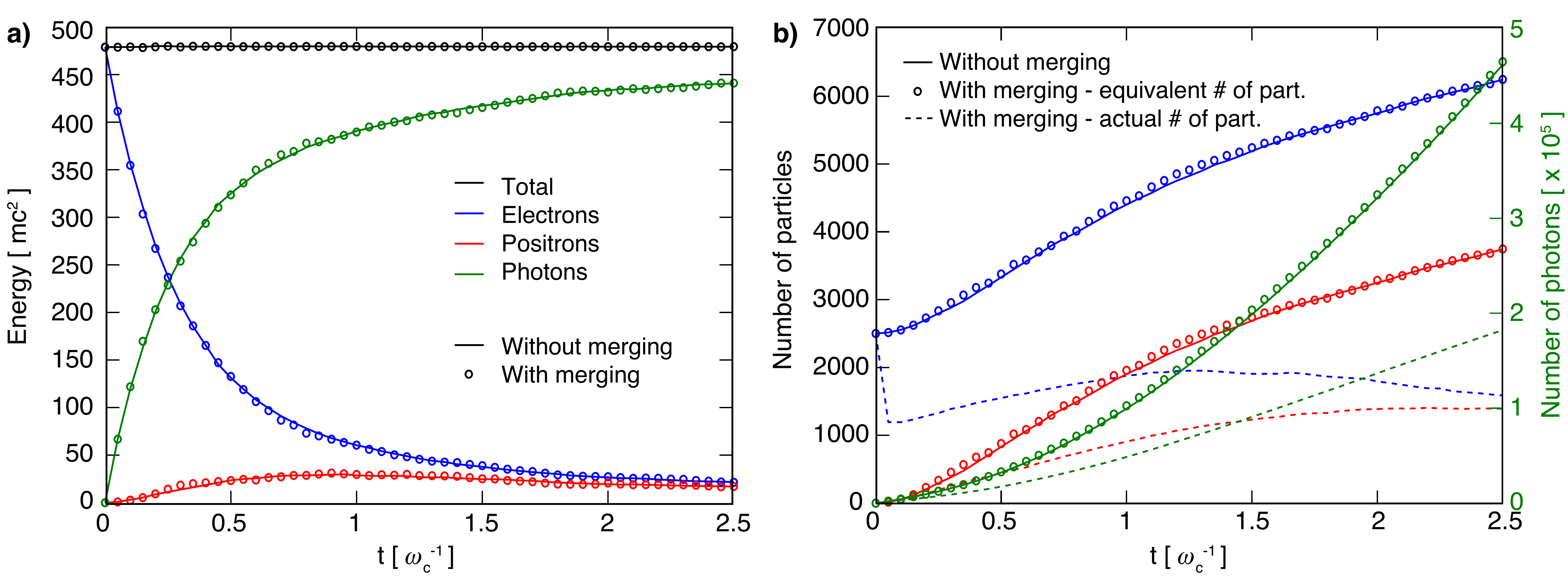}
\caption{ Magnetic showers. a) Energy conservation. The energy transfers from electrons to photons and positrons, but the total energy remains unchanged. b) Number of particles in the simulation without merging, with merging and equivalent number of particles with merging. } 
\label{merge_shower}
\end{figure}

In order to simulate the creation of electron-positron pairs, we have added a QED module which allows real photon emission from an electron or a positron and decay of the photons into pairs (Breit-Wheeler process). The implementation of such a module and the associated differential probabilities used can be found in \cite{NikishovRitus,pair_rate1,Ritus_thesis,pair_rate2,pair_rate3,Erber}. The specific points of the implementation in OSIRIS are presented elsewhere. One of the challenges of QED-PIC simulations is  the emergence of a vast number of particles (hard photons, electrons and positrons) that makes the simulations rather demanding. Among the QED scenarios where this is readily visible are magnetic showers.

Magnetic showers consist in an avalanche of electron-positron pairs produced by the decay of energetic photons in an ultra intense magnetic field \cite{Akhiezer,Anguelov_Vankov}. The setup we have chosen is a simple scenario identical as in ref. \cite{Gremillet} where a monoenergetic relativistic electron beam propagates initially perpendicularly to a uniform magnetic field. In a purely classical case the beam would describe a circular orbit. However due to the extreme magnitude of the magnetic field (few percent of the Schwinger field), the electrons emit hard photons (through quantum synchrotron radiation) that eventually make the beam slow down and thus spiral down. The photons emitted in the plane perpendicular to the magnetic field decay into pairs which in turn radiate new photons. The process is repeated until the initial energy of the electron beam is fully converted into an electron-positron-photon plasma. The growth of such an avalanche is not exponential since at every step of the process the new pairs created cannot get further energy in the magnetic field and have thus a lower energy than the initial electron (or positron) they are originating from. The time evolution of the number of particles and their corresponding energy in a magnetic shower is depicted on Fig. \ref{merge_shower}. Figure \ref{merge_shower} a) shows that the energy transfer that occurs between the species (the initial energy of the electron beam is converted into photons and positrons as well as lower energy electrons) is well reproduced when the simulation is carried out with the merging algorithm. The number of particles of every species as a function of time is shown in Fig. \ref{merge_shower} b). In order to compare both simulations, we also show the equivalent number of particles for the merged simulation, represented by circles in Fig. \ref{merge_shower} b), that we calculate by summing the weights of all the particles. The real number of simulation particles of every species with merging turned on is represented by the dashed lines.  We observe that these weighted particles, created during the merging process, mimic well the physics of the magnetic showers since the number of equivalent particles agrees at any time with the number of particles obtained in the simulation performed without merging. In this example of magnetic shower, the merged particles represent on average three to five non-merged particles leading to a simulation speed-up of 2.5 (see Table \ref{table:merge_param}).

\subsection{QED cascades}

\begin{figure}[t!]
\centering
\includegraphics[width=1.0\textwidth]{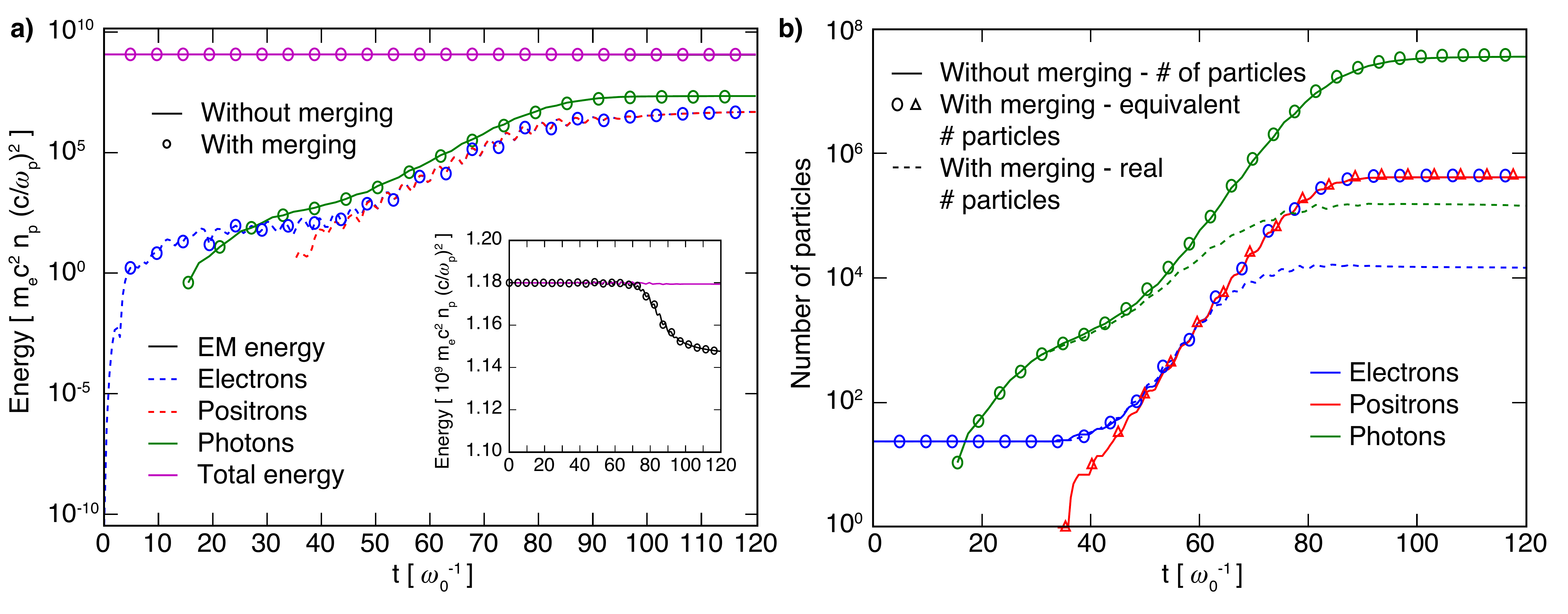}
\caption{ Cascade - a) Energy conservation. The energy transfers from the lasers to the electrons and positrons as they get accelerated; later, some of this energy is converted to photons through radiation emission. The inset shows a small level of laser absorption.  b) Number of particles with and without merging, and equivalent number of particles with merging. } 
\label{merge_casc}
\end{figure}

\begin{figure}[t!]
\centering
\includegraphics[width=1.0\textwidth]{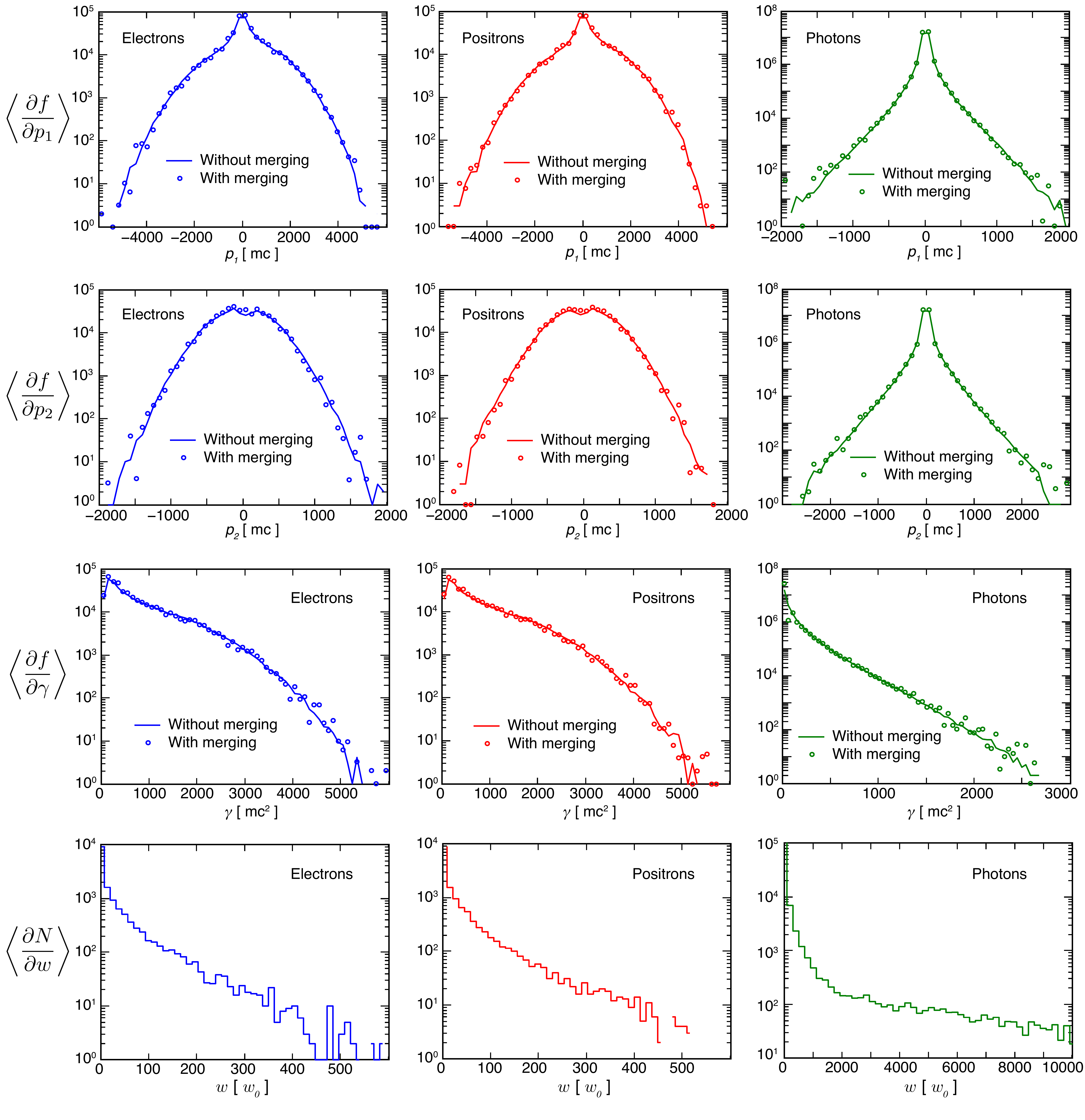}
\caption{ Cascade (top to bottom) - particle distributions in longitudinal momenta, transverse momenta and energy; weight distribution as compared with the initial weight} 
\label{dist_casc}
\end{figure}

The QED cascades are characterised by the creation of a pair plasma in a strong laser field \cite{Nerush, model1bell, Fedotov_cascade}. They differ from the usual pair avalanches in the fact that the newly created electrons and positrons are accelerated in the laser field and produce a new generation of photons and pairs similar to their ancestors. The process is then self-similar at every stage and one can expect an exponential growth of the number of pairs. We have a set-up similar to the one proposed by \cite{model1bell} and that has been first simulated by \cite{Nerush}. The cascade is seeded by a few immobile electrons that are located in the central position between two counter-propagating laser pulses. The two laser pulses have temporal Gaussian envelopes with a duration of 60 fs each and the focal spot size is about 10 $\mathrm{\mu m}$ at the location of the electron cloud initially placed at the centre of the simulation box. The additional parameters of the simulation can be found in Table \ref{table:merge_param}. The noticeable difference, in comparison with the magnetic showers, is the exponential growth rate of the number of pairs in the laser field zone. The time evolution of the number and the energy of the produced pairs and photons are identical with and without merging as we can see in Fig. \ref{merge_casc}. There is also no noticeable difference in energy conservation between the two cases.

Even if the main point of this study is not to dwell on the physics of QED cascades, it is also worth mentioning that the self-consistent created pair plasma reaches the relativistic critical density which in turn depletes, due to laser absorption (converted into thermal energy), a fraction of the initial electromagnetic energy. The inset of Fig. \ref{merge_casc} a) shows a laser depletion of 3\%. The simulation parameters were chosen to keep this value low in order to allow a direct comparison between simulations with and without merging, since the exponentially growing number of particles that results from this scenario would eventually cause the non-merged simulation to run out of memory. Whereas the examples we have discussed previously also showed us the good reproducibility of usual setups with merging, it is truly in scenarios such as QED cascades that the advantage of the algorithm becomes apparent as one notices that the number of pairs/photons is kept low (a factor of 1000 lower in the simulation with merging) in comparison with the standard run, see Fig. \ref{merge_casc} b). This considerable reduction of the number of PIC-particles does more than compensate the overhead time due to the merging process and leads to a simulation speed-up of 22 in this particular run. If one envisages to perform QED-PIC simulations that aimed to augment by several orders of magnitude the number of pairs/photons created, it is clear that a standard QED-PIC code would be unable to accomplish such a task and that the only way to perform these simulations is to rely on a merging algorithm. 

The nine top insets of Fig. \ref{dist_casc} compare the distribution functions for the merged and non-merged simulations, showing the space averaged momentum $\langle\partial f/\partial p\rangle$ and energy $\langle\partial f/\partial \gamma\rangle$ distributions of the electrons, positrons and photons at time $\omega_0t=100$ corresponding to the end of the cascade. Despite some small fluctuations in the tail of the distributions, the merged pairs/photons follow the same distributions as the non-merged ones. We have additionally plotted in Fig. \ref{dist_casc} the total weight distributions $\langle\partial N/\partial w\rangle$ of each species (electrons, positrons and photons). The pioneering works of \cite{NikishovRitus,pair_rate1,Ritus_thesis,pair_rate2,pair_rate3,Erber} tell us that photon emission is a more probable process than pair creation. It results in a higher number of photons merged than pairs and consequently the photon population is distributed over higher weights. In this simulation, the distributions of pairs spread up to $w\sim500~w_0$ (initially the electrons seeding the cascade had the weight $w_0$) whereas the weight of photons can reach $w\sim 10^4~w_0$. We observe that, if required, it is also conceivable to establish  an upper bound for the weight of the merged particles. Notwithstanding, all of the weight distributions display similar shapes: a bulk localised at small weights and a lower number  of particles in an exponential tail at high weights. It is an additional indication that the dynamical merging does not jeopardise the PIC statistics, as the particles outside of the vastly occupied regions of 6D phasespace retain their original weights.

\section{Conclusions}\label{sec:conclusions}
In summary, the particle merging algorithm for PIC simulations that has been implemented and tested conserves locally energy, momentum and charge by a detailed resampling of the 6D phase space. This particle merging algorithm naturally favors resampling the bulk of the particle distribution, and leaving the areas with a small statistical sample intact. When using this scheme, one should be aware of the typical length scales associated with the physics of the simulation and choose the size of the merging cells accordingly. 

We have studied the influence of merging on the simulation results in classical and QED scenarios, by comparing the full-PIC simulations with and without merging of particles. The presented scheme is found to be very successful in reproducing results both for linear and nonlinear plasma processes. We have also provided an estimate of the expected merging rate, that could help in the design and planning of the simulations. 

This algorithm will allow for significant speedups whenever the number of numerical particles in the simulation grows significantly (e.g. due to ionisation or pair-creation mechanisms). For instance, a huge speedup of QED cascading simulations has been verified and has enabled us to simulate problems in timescales otherwise impossible due to the exponential growth rate of the number of particles. Local conservation of energy, momentum and charge minimises the effect of resampling on the underlying physics, so it can be used also to improve the load balance of many other PIC simulations that encounter strong particle grouping both in real and in momentum space. More importantly, with marcroparticle merging algorithm it is now possible to explore problems that would otherwise not be accessible due to memory limitations.

\section*{Acknowledgement}
M. Vranic and T. Grismayer contributed equally to this work. This work was partially supported by the European Research Council (ERC-2010-AdG Grant 267841) and FCT (Portugal) grants FCT/IF/01780/2013, SFRH/BD/62137/2009. We would like to acknowledge the assistance of high performance computing resources (Tier-0) provided by PRACE on JuQUEEN and SuperMUC based in Germany. Simulations were performed at the IST cluster (Lisbon, Portugal), JuQUEEN and SuperMUC (Germany).






\bibliographystyle{elsarticle-num}
\bibliography{merging_final.bbl}







\end{document}